\documentclass[aps,prb,twocolumn,showpacs,groupedaddress,amsmath,amssymb]{revtex4}
\usepackage[dvips,usenames]{color}
\usepackage[dvips]{pstcol} 
\usepackage{nicefrac}		
\usepackage{bbm} 				
\usepackage{graphicx}
\usepackage{hyperref}
\usepackage[all]{hypcap}

\def \be{	\begin{displaymath}	}
\def \ee{	\end{displaymath}		}          

\def \ben{\begin{equation}  }
\def \een{\end{equation}    }            

\def \bea{\begin{eqnarray*}	}            
\def \eea{\end{eqnarray*}		}

\def \bean{\begin{eqnarray}	}            
\def \eean{\end{eqnarray}		}

\def \nn{\nonumber}

\def \Li{\mbox{Li}}

\def \Re{ \mbox{Re} }

\def \tr{ \mbox{{\rm tr}}}
\def \e{ {\rm e}}

\def \eps{\varepsilon}

\def \cl#1{ {\cal #1} }               
\def \b#1{ \mathbf{#1} }             

\def \bk{\b k}

\def \binomial#1#2{ \left( {#1} \atop {#2} \right) }

\def \Ref#1{(\ref{#1})}

\def \inv{ ^{-1} }
\def \dag{^\dagger}

\def \invb#1 { \frac{1}{#1} }
\def \bra#1{\langle #1 |}             
\def \ket#1{| #1 \rangle }            
\def \pro#1{ \ket{ #1 }\!\bra{ #1 } } 

\def \sp#1{ \langle #1 \rangle }               

\def \fr#1#2{ \frac{#1}{#2} }

\def \lite#1{[\cite{#1}]} 

\newcommand{\diff}[1]{\! \mathrm{d}#1 \hspace{0.2em} }

\pacs{02.50.Ga, 03.65.Yz, 03.67.-a, 03.67.Mn, 03.67.Pp, 75.10.Pq, 89.70.+c}

\begin{document}
\bibliographystyle{h-physrev3}	
\input epsf

\title{Decoherence of encoded quantum registers}
\author{Stefan Borghoff and Rochus Klesse}
\affiliation{Universit\"at zu K\"oln, Institut f\"ur Theoretische Physik, Z\"ulpicher Str. 77, D-50937 K\"oln, Germany}
\date{August 28, 2007}
\begin{abstract}
In order to eliminate disturbing effects of decoherence, encoding of 
quantum information in decoherence-free subspaces has been suggested.
We analyze the benefits of this concept for a quantum register that 
is realized in a spin chain in contact with a common bosonic bath. 
Within a dissipation-less model, we provide explicit analytical
results for the average fidelity of plain and encoded quantum
registers. For the investigation of dissipative spin-boson couplings,
we employ a master equation of Bloch-Redfield type.  
\end{abstract}
\maketitle

\section{Introduction}\label{sec-introduction}
The main obstacle in utilizing the remarkable computational power
of quantum systems \cite{feynman82,deutsch85,shor94} is
the omnipresent and fundamental phenomenon of decoherence
\cite{giulini96,zurek03}. 
While this insight cast significant doubts about the 
idea of large-scale quantum computation\cite{unruh94,chuang95}, it also initiated extensive research on decoherence in quantum information systems,
and, beyond that, it led to the development of quantum error-correcting\cite{shor95,calderbank96,steane96} and -avoiding\cite{palma97,zanardi97,lidar98} methods.
The latter ones, on which we focus in the present work, make
use of possible symmetries in the interaction of, say, a quantum
register and its surrounding environment. The idea is to encode
quantum information in those register states that are protected by
symmetry against the decohering interaction. 
Inasmuch as the symmetry is satisfied these states span a decoherence-free subspace in the register's Hilbert space. 

Almost necessarily physical realizations of this concept  
will have to rely on symmetries that hold only to some
approximation. Encoding in subspaces that respect these symmetries
can then provide only partial protection against decoherence, to an
extent that will depend on the actual realization. 
The present work addresses this problem for the generic situation of
a quantum register consisting of (effective) spin-$\nicefrac{1}{2}$
particles in contact with a common bosonic bath. 

In the main (Secs.~\ref{sec-decoherence} and \ref{sec-fidelity}), we
describe this system by the dissipation-less spin-boson model of Palma
et al.~\cite{palma97} which has been frequently used in similar contexts
\cite{duan98,reina02,klesse05,doll07}. For
simplicity, we assume the spins to be arranged in a linear chain with
inter-spin distance $a$. Furthermore, we will use a three-dimensional
bosonic bath with an ohmic coupling density of states.
In the limit of vanishing distance $a$ the
model exhibits a highly symmetric spin-boson interaction, allowing the
construction of decoherence-free subspaces. Specifically, we consider
subspaces that correspond to encoded quantum registers in which
logical qubits are encoded in locally grouped physical qubits (spins)
(cf.~Sec.~\ref{subsec-encoded}).

The main task is to analyze the decoherence which will appear in
these encoded quantum registers when the distance $a$ assumes finite
values. 
In Sec.~\ref{sec-fidelity} we quantify the decoherence of 
encoded registers (with finite distance $a$) as well as of plain
registers by means of the average register fidelity. 
Sec.~\ref{sec-dissipative} is devoted to the effect of dissipative
spin-boson couplings. 

For a summary of our results we refer to the self-contained
presentation in Sec.~\ref{sec-summary}.

\section{Decoherence of quantum registers}\label{sec-decoherence}

\subsection{Physical quantum register}\label{subsec-plain}
A physical $n$-qubit register may consist of $n$ (effective)
spin-$\nicefrac{1}{2}$ particles located 
at sites $\b r_0,\b r_2, \dots,\b r_{n-1}$ of a one-dimensional
lattice with lattice constant $a$ and of finite length $L=a(n-1)$.
A homogeneous (effective) magnetic field in $z$-direction may lead to a Zeeman
energy splitting $\eps$. The corresponding register Hamiltonian is
\be
H_R = \fr{\eps}{2} \sum_{l=0}^{n-1} Z_{l}\:,
\ee
where $Z_l$ denotes the Pauli $\sigma_z$-operator for spin $l$. 
The register is supposed to be in contact with a thermal bath of
three-dimensional bosons described by the Hamiltonian 
\be
H_B = \sum_{\b k} \hbar \omega_{\b k} b_{\b k}\dag b_{\b k}\:,
\ee
where $b\dag_{\b k}$ and $b_{\b k}$ are creation and annihilation operators of bosonic modes with linear dispersion $\omega_{\b k} = c |\b k|$. 
We assume a linear and local spin-boson interaction via
$Z_{l} b_\bk\dag$ and $Z_{l} b_\bk$ operators. 
The corresponding coupling constant $g_{l,\bk}$ 
will acquire a phase $e^{i\bk\cdot\b   r_l}$, reflecting the
wave-like character of the bosonic modes. Apart from this 
phase the interactions may be isotropic and identical for each
spin. Thus, $g_{l,\bk} = g_{|\bk|} e^{i\bk\cdot\b r_l}$, resulting in 
an interaction Hamiltonian
\be 
H_{RB} = \sum_l Z_l B(\b r_l)\:,
\ee
where $B(\b r)$ is the hermitian bosonic field operator
\be
B(\b r) = \sum_\bk g_{|\bk|} e^{-i \bk \cdot {\b r}} b_\bk\dag \: + \:
H.c. \:.
\ee
As customary, we describe the strength of the spin-boson coupling 
by a spectral density
$$
J(\omega) = \sum_\bk \delta(\omega_\bk - \omega) |g^{{}}_{|\bk|}|^2 
\: \equiv \: \alpha \omega^s
\e^{-\omega/\Omega}\:.
$$
It is characterized by a cut-off frequency $\Omega$, a constant $\alpha$ of
appropriate dimension, and a non-negative spectral parameter $s$
\cite{weiss99,breuer02}.
Since the spins interact with the bosons only via the energy conserving 
$Z_l$ operators, the model shows no dissipation but pure
decoherence. This restriction makes the model analytically 
manageable.\cite{palma97,reina02,breuer02,doll07} (Effects of additional dissipative couplings will be
discussed in Sec.\ref{sec-dissipative}.)

\subsubsection{Decoherence of a physical quantum register}
\label{subsubsec-decoherence-physical} 
The register may be used to store quantum information 
in form of a state $\rho_0$ in which it is initially prepared. 
In general, $\rho_0$ is subjected to a non-unitary dynamics
originating from the system's own dynamics and its coupling to the
bosonic bath. 
Assuming that the bath is initially in a thermal state $\rho_B$, the
total initial state is  $\varrho_0 = \rho_0 \otimes \rho_B$. During
some time period $t$ this state will evolve unitarily according to
$i\hbar \dot\varrho = [H_R + H_B + H_{RB}, \varrho]$ towards a final state
$\varrho_t$. Its partial trace with respect to the bosonic modes 
yields the reduced density operator $\rho_t = \tr_B \varrho_t$, which describes
the final register state.
This procedure defines a quantum operation\cite{nielsen00,breuer02} $\cl E$ on the register by 
\be
\rho_0 \mapsto \cl E(\rho_0) := \rho_t\:.
\ee
The work of several authors\cite{palma97,duan98,bacon99,reina02,breuer02,zurek03,doll07} established the operation $\cl E$ to be of the
form 
\be
\cl E (\rho) = \cl U \circ \cl N (\rho)\:,
\ee
where $\cl U$ is a purely unitary operation,
and $\cl N$ is a non-unitary, completely positive map that can be written as
\ben\label{N-operation}
\cl N (\rho) = \sum_{\mu\nu\in \b Z_2^n } e^{-D_{\mu\nu}} \pro{\mu}
\rho \pro{\nu}\:. 
\een
Here, the double summation extends over all register eigenstates $\ket
\nu$ which we label in the usual way by $n$-bit sequences 
$\nu  \in \{ 0,1 \}^n \equiv \b Z_2^n$
according to $Z_l \ket \nu ~ = ~ (-1)^{\nu_l} \ket \nu$. The decoherence
coefficients $D_{\mu\nu}$ are
\begin{equation}
\label{decoherence-coefficients}
D_{\mu\nu} = \sum_{lm=0}^{n-1} (\mu_l-\nu_l)(\mu_m-\nu_m) K(|\b r_l - \b
r_m|, t)\:,
\end{equation}
with a distance and time dependent decoherence function
\ben\label{correlator-integral}
K(|\b r|,t) = 4 \Re \int\limits_0^t \diff{t'} \int\limits_0^{t'} \diff{t''} \sp{ B(\b r, t'')B(\b 0,0) }_T\: .
\een
Here $\sp{\dots}_T$ denotes the thermal average over the bosonic system at temperature $T$,
and $B(\b r, t)$ is the bosonic field operator in interaction picture, i.e.
\begin{equation}
	\label{bosonic-field}
	B(\b r, t) = \sum_\bk g_{|\bk|} e^{-i ({\bk} \cdot {\b r}_l - \omega_{|\bk|} t)} b_\bk\dag \: + \: h.c. \:.
\end{equation}

The unitary part $\cl U$ of $\cl E$ originates from the registers's own
dynamics but also includes the Lamb-shift caused by the bosonic
bath. In principle, this part of the evolution $\cl E$ can be reversed  
and therefore is not of major concern. In contrast to that, the
operation $\cl N$ gives rise to decoherence  and so seriously affects
the register in an irreversibly manner. 
Clearly the main attention has to be paid to $\cl N$.
Therefore, it will be in the focus of our investigation. 

\subsubsection{Ohmic decoherence function}\label{subsubsec-ohmic}
In order to make $\cl N$ more explicit, we have to determine the decoherence function Eq.~(\ref{correlator-integral}).
For the sake of simplicity, we restrict ourself to an ohmic spectral function  $J(\omega) = \alpha \omega e^{-\omega/\Omega}$. 
Then, determining the correlator $\sp{ B(\b r, t'')B(\b 0,0) }_T $ by standard methods and passing the continuum limit for the bosonic modes, the decoherence function
\Ref{correlator-integral} becomes
\ben\label{decoherence-function}
K(r,t) =  \alpha \int\limits_0^\infty \diff{\omega}  \fr{1-\cos \omega t}{\omega} \coth\left(\fr{\omega}{2 T}\right) \fr{\sin\omega r}{\omega r}e^{-\omega/\Omega} \:.
\een
(Henceforth we use units in which $c=1$, $\hbar=1$, and $k_B=1$.)
This integral can be better dealt with if we  
distinguish between the case of strictly vanishing distance $r$ and the
case of a finite distance that is large compared to the cut-off wavelength $\sim 1/\Omega$,
which we assume to be the smallest scale in the problem. 

For vanishing distance we obtain 
\ben\label{Knull}
K(0, t) = 2 \alpha \ln \left|\fr{ \Gamma(T/\Omega)}{\Gamma(T/\Omega
    - i t T)}\right| \: - \: \fr{\alpha}{2} \ln(1+t^2 \Omega^2)\:.
\een
At small times $t \ll 1/T$ this is in good approximation 
\ben\label{Knull-small}
K(0, t) \:\simeq \:\fr{\alpha}{2} \ln(1+t^2\Omega^2)\:,
\een
whereas for large times $t\gg 1/T$ we have
\ben\label{Knull-large}
K(0,t) \:\simeq\: \alpha \pi T t + \alpha \ln \fr{\Omega}{2 \pi T}\:.
\een
Physically, $K(0,t)$ determines the decoherence of a single spin, as it
is seen by Eqs.~\Ref{decoherence-coefficients} and \Ref{N-operation}
for $n=1$, according to which 
\be
\rho_{01}(t) = e^{-K(0,t)} \rho_{01}(0)\:.
\ee
Asymptotically, the single-spin decoherence decays exponentially with
a rate $\gamma = \alpha \pi T$, by Eq.~\Ref{Knull-large}.

For finite $r \gg 1/\Omega$ the oscillations of the spherical Bessel 
function $\sin (\omega r)/\omega r$ 
damp the integrand in Eq.~\Ref{decoherence-function} more effectively
than the regular cut-off 
$\exp(-\omega/\Omega)$. This allows us to take the limit $\Omega \to
\infty$. The resulting integral can be solved by contour integration,
which finally leads us to 
\bean
K(r,t) &=& \alpha \pi T (t - \fr{r}{2} +  \fr{1}{12 T^2 r}) \: + \: \label{Kexact1}\\
& &  \fr{\alpha}{4\pi T r}\left[ f_T(t+r) -f_T(t-r)  -
  2f_T(r)\right]\:, \nn
\eean
for $r<t$, and to 
\bean
K(r,t) &=& \alpha \pi T \:\fr{t^2}{2r} \:\:+\label{Kexact2}\\
& &  \fr{\alpha}{4 \pi T r}\left[f_T(t+r) +
  f_T(r-t) - 2 f_T(r) \right]\:, \nn
\eean
for $r>t$. 
For convenience, we introduced  a temperature dependent function
$f_T(t):=\Li_2(e^{-2\pi T
  t})$, where $\Li_2(x) ~ = ~ \sum_{j=1}^\infty x^j/j^2$ is the
dilogarithm of $x$. 

\begin{figure}
\vspace{0.5cm}
  \begin{center}
    \epsfxsize=7.0cm
    \epsffile{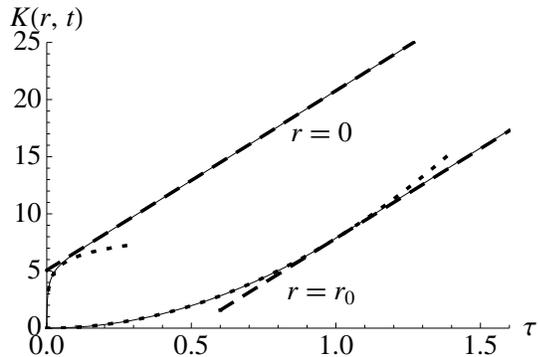}
    \vspace{0.3cm}
    \caption{
			Decoherence functions $K(0,t)$ and $K(r_0,t)$ as a function of dimensionless time $\tau = t / r_0$ at temperature $T = 5 / r_0$.
			The cut-off energy is $\Omega = 10^3 T$.
			The dotted and dashed curves are short- and long-time approximation, respectively.
    	\label{fig-decoherence}
		}
  \end{center}
\end{figure}

In the large temperature regime characterized by $t,r,|t-r| \gg 1/T$,
the $f_T(\cdot)$-terms in Eqs.~\Ref{Kexact1} and \Ref{Kexact2} become
exponentially suppressed, and so, additionally omitting an $ \alpha
\pi /24 T r $ term, 
\bean
K(r,t) &\simeq& \alpha \pi T \left( t - \fr{r}{2} \right)\:,
\quad \mbox{for} \:r<t  \label{largeT1} \\ 
K(r,t) &\simeq& \alpha \pi T \:\fr{t^2}{2r}\:, \qquad\: \:\mbox{for} \:r>t \label{largeT2}
\eean
Fig.~\ref{fig-decoherence} shows $K(0,t)$ and $K(r_0,t)$ as a function
of dimensionless time $\tau = t / r_0$ together with their approximations Eq.s~\Ref{Knull-small},\Ref{Knull-large},
\Ref{largeT1}, and \Ref{largeT2}, for $\Omega = 10^3 T$.

\subsection{Encoded quantum register}\label{subsec-encoded}
For vanishing lattice constant $a$ the locations of all spins of the quantum register introduced in \ref{subsec-plain} fall onto a single point $\b r_0$. 
This implies a highly symmetric spin-boson interaction
\ben\label{symmetric-hamiltonian}
H_{RB}^{(0)} = (\sum_l Z_l )\: B(\b r_0)
\een
that exactly annihilates all states with vanishing total spin-$z$
component. 
As a consequence, any linear subspace $\cl C$ of  
the register's state space $H_n$ that is spanned by such states is
not affected by the bosonic bath at all. It represents a decoherence-free
subspace \cite{palma97,zanardi97,lidar98}. 

At a finite lattice constant $a$ the former symmetry is absent and consequently $\cl C$ ceases to be decoherence-free. 
However, by reasons of continuity the decoherence of states in $\cl C$ will be still much lower than for arbitrary states as long as the lattice constant $a$ is not too large. 

Of course, the decoherence reduction at a finite lattice constant $a$ will
strongly vary for different choices of the subspace $\cl C$. 
Here we will investigate subspaces that result from encoding (logical)
qubits in local groups of physical qubits. This is supposed to be done
in a regular manner such that the resulting structure forms a regular
encoded quantum register. 

\begin{figure}[ht!]
  \centering
	\includegraphics[scale=0.6]{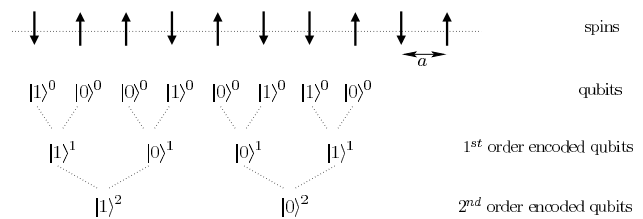}
  \caption{Spin array representing qubits and encoded logical qubits of $1^{st}$ and
    $2^{nd}$ order.}
  \label{fig-subspaces}
\end{figure}

To be specific, let us consider a one-dimensional physical $2n$-qubit
register $R_{2n}$ whose $2n$ spins $S_0, \dots, S_{2n-1}$ are grouped in $n$
pairs of neighboring spins as sketched in Fig.~\ref{fig-subspaces}.
The spin pair $S_{2i}S_{2i+1}$ has a four-dimensional Hilbert space
of which the two orthonormal states 
\ben\label{logical-kets}
\ket{0}_i^{1} := \ket{0}_{2i} \ket{1}_{2i+1}, \quad
\ket{1}_i^{1} := \ket{1}_{2i} \ket{0}_{2i+1}\: \quad
\een
are annihilated by the spin-boson interaction if $a=0$. 
In this case the subspace $C_i$ spanned by states $\ket{0}_i^1$ and $\ket{1}_i^1$ is decoherence-free.
We call $C_i$ the state space of the encoded
(logical) qubit $Q_i$, and we further define an encoded $n$-qubit
register $R_n^{1}$ to consist of the $n$ encoded qubits $Q_0, \dots
,Q_{n-1}$.  Its Hilbert space $H_n^1= C_0\otimes \dots \otimes C_{n-1}$
is by construction decoherence-free with respect to $H_{RB}^{(0)}$.
We will denote a state $\rho$ as a state of the encoded register $R_n^1$
if the support of $\rho$ lies entirely in $H_n^1$.

Clearly, pairing up remote spins instead of adjacent ones would lead
to encoded qubits that will be more sensitive to an increasing lattice
constant $a$. We will therefore exclude this possibility from our considerations.
Instead, one may speculate that the protection against decoherence
improves if we iterate the pairing in order to built 
encoded qubits and registers of higher order (cf.~Fig.~\ref{fig-subspaces}). 

More precisely, we recursively define encoded qubits of order
$\chi=1,2,\dots$ by their logical 
states
\be
\ket{0}_i^{\chi} := \ket{0}_{2i}^{\chi-1} \ket{1}_{2i+1}^{\chi-1}, \quad
\ket{1}_i^{\chi} := \ket{1}_{2i}^{\chi-1} \ket{0}_{2i+1}^{\chi-1}\:, \quad
\ee
where $0^{th}$ order states are identified with plain spin states
$\ket{0}^0_i = \ket{0}_i$ and $\ket{1}^0_i = \ket{1}_i$ of spin $S_i$.
An encoded $n$-qubit register $R_n^\chi$ of order $\chi$ is then built
from the encoded qubits of a $2n$-qubit register $R_{2n}^{\chi-1}$ of
order $\chi-1$. 
A state $\rho_\chi$ of the encoded register $R_n^\chi$ is by definition a
state whose support lies in $H_n^\chi$.

\subsubsection{Decoherence of encoded quantum register}\label{subsubsec-decoherence-of-encoded}
How will the decohering operation $\cl N$ affect the encoded registers which we have just introduced?
First we observe that a state $\rho_\chi$ of an encoded register  $R_n^\chi$ remains a state of $R_n^\chi$ under $\cl N$, simply because  the spin-boson interaction $H_{RB}$ does not flip spins. 
Moreover,  in App.~\ref{app-decoherence-of-encoded} we show the
operation $\cl N$ on an encoded register $R_n^\chi$ to be formally given
again by Eqs.~\Ref{N-operation} and \Ref{decoherence-coefficients}.  
What changes  is the decoherence function $K(|\b r_l - \b r_m|,t)$, 
which has to be replaced by an effective decoherence function
$K^\chi_{|l-m|}(t)$, and the summation, which now extends over
logical register states $\ket{\mu}^\chi, \ket{\nu}^\chi $ given by
\be
\ket{\mu}^\chi = \ket{\mu_0}_0^\chi \dots \ket{\mu_{n-1}}_{n-1}^\chi 
\:, \quad \mu \in \b Z_2^n\:.
\ee
Explicitly, for a state $\rho_\chi$ of an encoded register $R_n^\chi$
we have 
\ben\label{chi-N-operation} 
\cl N(\rho_\chi) = 
\sum_{\mu\nu \in \b Z_2^n} e^{-D_{\mu\nu}^\chi}
\ket{\mu}^{\!\chi}\!\bra{\mu} \rho_\chi \ket{\nu}^{\!\chi}\!\bra{\mu}\:,
\een
with effective decoherence coefficients 
\ben\label{chi-decoherence}
D^\chi_{\mu\nu} = \sum_{lm=0}^{n-1}  (\mu_l -\nu_l)(\mu_m-\nu_m)
K_{|l-m|}^\chi(t)\:. 
\een
The effective decoherence functions $K^\chi_{l}(t)$ for $\chi \ge 1$
are recursively defined by
\ben\label{chi-decoherence-function}
K_{l}^\chi(t) = 2 K_{2l}^{\chi-1}(t)-K_{|2l-1|}^{\chi-1}(t) -
K_{2l+1}^{\chi-1}(t)\:,   
\een
with 
$
K^0_{l}(t) = K(la, t)
$.

\subsection{Discussion}\label{subsec-discussion}
The formal analogy of 
Eqs.~\Ref{chi-N-operation},\Ref{chi-decoherence}
and 
Eqs.~\Ref{N-operation} \Ref{decoherence-coefficients} 
allows for a first comparison of the decoherence in encoded and plain
quantum registers by simply comparing the
corresponding decoherence functions given by
Eq.~\Ref{chi-decoherence-function} and 
Eq.~\Ref{decoherence-function}. 

\begin{figure}
\vspace{0.5cm}
  \begin{center}
    \epsfxsize=7.0cm
    \epsffile{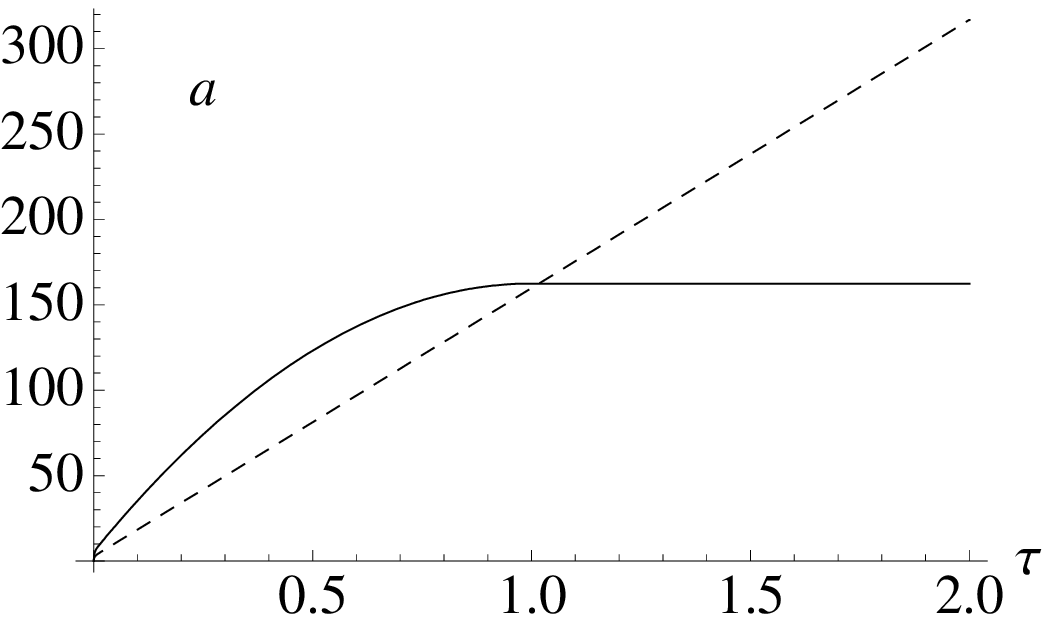}
    \epsfxsize=7.0cm
    \epsffile{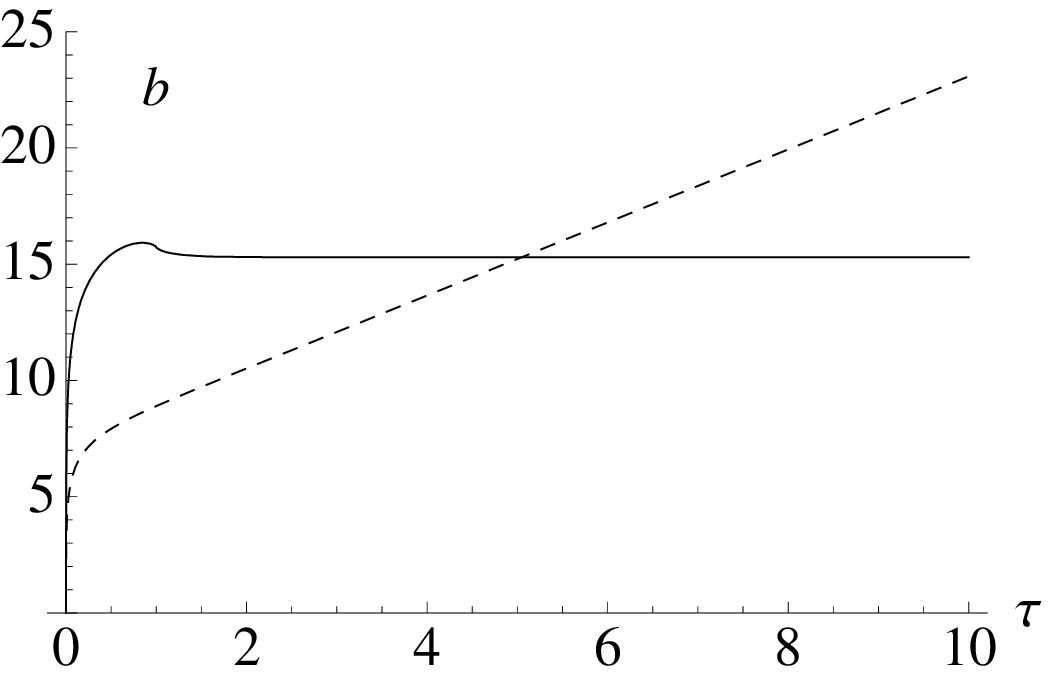}
    \vspace{0.3cm}
    \caption{
			Decoherence functions $K(0,t)$ (dashed) and $K_0^1(t)$ (solid) as function of dimensionless time $\tau = t / a$  
			at high (a) and low (b) temperatures with cut-off $\Omega = 10^3 / a$, and $T=10 / a$ in (a) and $T=0.1 / a$ in (b). 
    \label{fig-Knull}}
  \end{center}
\end{figure}

We begin with the $1^{st}$ order decoherence function $K_0^1(t)$, which
describes the effective decoherence of a single encoded qubit
$R_1^1$.
According to Eq.~\Ref{chi-decoherence-function} 
\be
K_0^1(t) = 2 ( K(0,t) - K(a,t) )\:.
\ee
In the high temperature regime $t,a\gg 1/T$ we may use
approximations \Ref{Knull-large}, \Ref{largeT1}, and \Ref{largeT2} to 
derive
\be
K_0^1(t) = 2 \alpha \ln \fr{\Omega}{2 \pi T} \: + 
\left\{
\begin{array}{lcl}
\alpha \pi T a & : & t >a \\
& & \\
\alpha \pi T( 2t - \fr{t^2}{a}) &: & t \le a 
\end{array}
\right.
\ee
The effective decoherence function $K_0^1(t)$ increases twice as fast
with time as $K(0,t)$ for small times, but quickly saturates to a
constant value at time $t\simeq a$ (cf.~Fig.~\ref{fig-Knull}a). 
Qualitatively, this remains to be also true at lower temperatues.
For $a \ll 1/T \ll t$ we can extract from relation \Ref{Knull-large}
and the exact expression \Ref{Kexact1} an asymptotic value
\be
K_0^1(\infty) \simeq 2 \alpha \ln \fr{\Omega a}{e}\:
\ee
that is reached again at $t \simeq a$ (cf.~Fig.~\ref{fig-Knull}b). 

We conclude that for any finite distance $a$ the coherence
$e^{-K_0^1(t)}$ of an encoded qubit approaches a finite asymptotic
value at times $t \gtrsim a$. 
In the long-time limit the coherence of the encoded qubit will 
therefore largely exceed the exponentially
decaying coherence $e^{-K(0,t)} = (2\pi T/\Omega)^\alpha e^{-\alpha
\pi T t}$ of a plain qubit.
At short times, however, the encoded qubit performs worse than the
plain qubit. 
The crossover time $t_c$ can be easily determined to be
$t_c \simeq a$ in the high temperature regime $T\gg 1/a$, and $t_c
\simeq \fr{1}{\pi T} \ln(2 \pi a^2 \Omega T/e^2) \gg a $ in the low temperature
limit $T \ll 1/a$. 
Thus, whether it is beneficial to encode or not also depends on the
time period over which the qubit is supposed to store information. 
Here, it is important to observe that with lowering the temperature $T$
one eventually reaches the low temperature regime where 
the crossover time $t_c$ increases with $1/T$.

Do things further improve when one goes to higher-order encoded
qubits? Interestingly, this is not the case, for the reason that 
the $1^{st}$ order qubits of an encoded register $R_n^1$ are already
essentially decoupled (see below), and pairing up these independent
qubits to higher-order qubits would not further reduce their effective
decoherence. 
The decoupling of the $1^{st}$ order-qubits is seen  
from the effective decoherence functions $K_l^1(t)$ for $l\ge1$. By
Eq.~\Ref{chi-decoherence-function} we find
\be
K_{l}^1(t) = 2 K(2la,t) - K( 2la-a,t) - K( 2la +a,t)\:.
\ee
In the high temperature regime this predicts by Eq.~\Ref{largeT1}
actually a vanishing $K_l^1(t)$ for times $t>2la + a$. 
More precisely, for $l \geq 1$
\be
K_{l}^1(t) = O \left( \fr{\pi}{24 T l a} \right) \: \ll \: K_l^0(t) \simeq  \alpha \pi T \left( t- \fr{la}{2} \right)\: .
\ee 

\begin{figure}
\vspace{0.5cm}
  \begin{center}
    \epsfxsize=7.0cm
    \epsffile{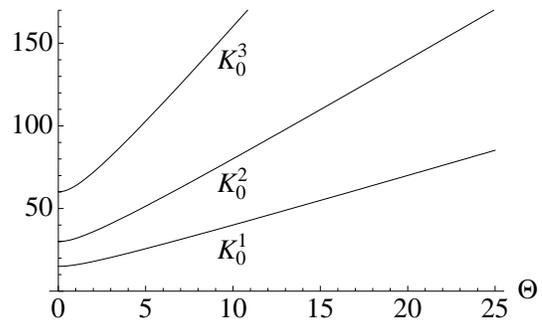}
    \vspace{0.3cm}
    \caption{
    	\label{fig-Knull-T}
			Asymptotic value of the decoherence functions $K_0^{\chi}(\infty)$ as function of dimensionless temperature $\Theta = T a$ for $\chi=1,2,3$ and cut-off $\Omega = 5 \cdot 10^3 / a$. 
		}
  \end{center}
\end{figure}

Alternatively, we can also directly calculate the zero-distance
decoherence functions $K_0^\chi(t)$ according to  relation
\Ref{chi-decoherence-function}.  
In the high
temperature regime we obtain in the long-time limit for $\chi \ge 1$
\be
K_0^\chi(\infty) = 2^{\chi-1} K_0^1(\infty)\:,
\ee
which obviously strongly increases with $\chi$. 
Qualitatively similar behaviour is found also at low temperatures,
where we used the exact result Eq.~\Ref{Kexact1} to numerically
determine $K_0^\chi(\infty)$. The results are plotted in Fig.~\ref{fig-Knull-T}

So far the discussion is restricted solely on a comparison of (effective) decoherence functions. 
While this suffices to characterize the decoherence of (encoded) qubits, it does not necessarily provide full insight in the performance of an entire (encoded) quantum register under the noise operation $\cl N$. 
The next section addresses this point by a systematic investigation of average register fidelities. 

Another important point which deserves further investigation is
dissipation. The above analysis is based to a large extent on exact
results that are available only for the dissipation-less spin-boson 
model. Therefore, it might be possible that essential features of
encoded registers -- particularly the 
saturation of the decoherence -- will not survive when dissipative
spin-boson couplings are taken into account. We will investigate this
problem in Sec.~\ref{sec-dissipative}.

\section{Average register fidelities}\label{sec-fidelity}

%

\subsection{Definition}\label{subsec-definition}
Let a physical or encoded quantum register $R_n$ with Hilbert space $H_n$
be exposed to some noise operation $\cl N$ (for instance, the one
studied above), under which an initial pure register state 
$\psi = \pro\psi$ evolves to a final $\cl
N(\psi)$. The channel fidelity of $\psi$ with respect to $\cl N$,   
\be
F(\psi, \cl N) := \sp{\psi| \cl N(\psi)|\psi}\:,
\ee
captures how well the state is preserved in this process\cite{nielsen00}. 
A quantum register will be suitable for information storage under the
noise $\cl N$ if, in average, the channel fidelity $F(\psi, \cl N)$
for register states is large.  
A reasonable figure of merit is therefore given by the 
average fidelity of register $R_n$ with respect to $\cl N$, 
\be
F := \fr{1}{\mbox{N}} \int\limits_{H_n} \! \diff{\psi} \: F(\psi,\cl N) \:,\quad \mbox{N} = \int\limits_{H_n} \! \diff{\psi} \: 1\:,
\ee
where the integrals extend over all pure code states $\psi$ with state vectors $\ket \psi \in H_n$ with respect to a unitary invariant measure. 

More precisely, we can express the average by an integration over 
the group $\b U(H_n)$ of unitaries on $H_n$ with 
the normalized Haar measure $\mu$,
\ben
F = \int\limits_{\b U(H_n)} \diff{\mu(U)} \: F(U \psi_0 U\dag,
\cl N)\label{def-average-fidelity} \:,
\een
where $\psi_0$ is any fixed pure register state.

For large qubit number $n$ the average register fidelity  is known to agree with the entanglement fidelity of $\cl N$ with respect to $H_n$, which plays a prominent r\^ole in quantum information theory.  

\subsection{General expressions}
\label{subsec-fidelity-plain}  

\subsubsection{Sum representation}\label{subsubsec-sum}
From now on we consider the noise operation $\cl N$ of the
dissipation-less spin-boson model which we introduced in Sec.~\ref{sec-decoherence}. 
Our aim is to derive an expression for the average fidelity 
for physical or encoded registers in terms of the decoherence
coefficients \Ref{chi-decoherence}.
Using representation  
\be
\cl N (\rho) = \sum_{\mu\nu =0}^{2^n-1} e^{-D_{\mu\nu}} \pro{\mu} \rho
\pro{\nu}
\ee
(we omit the order-index $\chi$, and identify $\b Z_2^n$ with integer numbers  $0,1, 2, \dots, n-1$) and relation \Ref{def-average-fidelity} with $\psi_0 ~ = ~ \pro{0}$ we immediately obtain
\be
F 
= \sum_{\mu\nu=0}^{2^n-1} e^{-D_{\mu\nu}}
\int\limits_{\b U(\cl H_n)}\!\! \diff{\mu(U)} \: |U_{0\mu}|^2 |U_{0\nu}|^2\:.
\ee
The integral can be calculated by standard methods\cite{pereyra82}, 
leading to   
\be 
\int\limits_{\b U(H_n)}\!\! \diff{\mu(U)} \: |U_{0\mu}|^2 |U_{0\nu}|^2
= \fr{1 \: + \: \delta_{\mu\nu}}{4^n} \: + \: O(2^{-6n})\:.
\ee

This outcome is easily understood once one recognizes
the integration as an average over all
$2^n$-dimensional complex unit vectors $u_0 \in H_n$ of which $U_{0\mu}$ and
$U_{0\nu}$ are the $\mu$th and $\nu$th component, respectively. 
For large dimension $2^n$ and $\mu\neq\nu$ the squared
absolute values $|U_{0\mu}|^2$ and $|U_{0\nu}|^2$ are nearly
independent gaussian variables $X_1$ and $X_2$ of mean
$\sp{X_1} = \sp{X_2} = 2^{-n}$, by the normalization of $u_0$.
For $\mu\neq\nu$ the integral over the product $|U_{0\mu}|^2
|U_{0\nu}|^2$ therefore amounts to the expectation value $\sp{X_1X_2}
= (2^{-n})^2 $. For $\mu=\nu$ we instead obtain the second moment 
$\sp{X_1^2} = 2 \sp{X_1}^2 = 2(2^{-n})^2$. 
The exact calculation reveals rather tiny corrections to these estimates
of order $2^{-6n}$, which we will neglect in the following. 
Then, by the last two equations we find 
\be
F = \fr{1}{4^n}  \sum_{\mu\nu=0}^{2^n-1} e^{-D_{\mu\nu} } (1
+\delta_{\mu\nu})\:.
\ee
Since the sum over the extra diagonal terms $e^{-D_{\mu\nu}} \delta_{\mu\nu}$
contributes at most $2^{-n}$ to the fidelity we can omit these terms
as well and thus are left with
\ben\label{sum-fidelity}
F = \fr{1}{4^n}  \sum_{\mu\nu=0}^{2^n-1} e^{-D_{\mu\nu} }\:.
\een
In general, the double summation over the exponentially large range
$0,\dots, 2^n-1$ makes a direct numerical or analytical
evaluation of this expression difficult. 
However, progress can always be made if 
we proceed similar as in Ref. \cite{klesse05} and employ a Hubbard-Stratonovich
transformation. This will factorize the double sum into $n$ trivial sums, at the
expense of an $n$-dimensional integration over auxiliary continuous
degrees of freedoms. 

\subsubsection{Integral representation} 
\label{subsubsec-integral}
We rewrite the decoherence coefficients
Eq.~\Ref{chi-decoherence} as 
\be
D_{\mu\nu} =  \b v\dag_{\mu\nu} \b K\: \b v_{\mu\nu}\:,
\ee
where we introduced real, $n$-dimensional vectors $\b v_{\mu\nu}$
with components
\be
(\b v_{\mu\nu} )_m = \mu_m - \nu_m\:, 
\ee
and a real and positive $n\times n$ decoherence matrix $\b K$ 
whose entries are determined by the (effective) decoherence functions
Eq.~\Ref{chi-decoherence-function}, 
\be
\b K_{lm} = K_{|l-m|}(t)\:. 
\ee
Then, with Gauss's identity 
\be
e^{-  \b v\dag_{\mu\nu} \b K\: \b v_{\mu\nu} } 
=  \int\! \fr{\diff{^n \b x}}{N}
\:e^{- \b x\dag \b K\inv \b x \: + \: 2 i \b
  v\dag_{\mu\nu} \b x} \:,
\ee
where  $N = \left(\pi^n \det \b K\right)^{\nicefrac{1}{2}}$, 
the average fidelity Eq.~\Ref{sum-fidelity} becomes
\be
F =  \int\! \fr{\diff{^n \b x}}{4^n N}
\:e^{- \b x\dag \b K\inv \b x }
\:\sum_{\mu\nu=0}^{2^n-1} 
e^{2 i \b v\dag_{\mu\nu} \b x} \:.
\ee
The sum is readily determined to be 
\be
 \sum_{\mu\nu=0}^{2^n-1} e^{2 i \b v\dag_{\mu\nu} \b x}
= \prod_{l=1}^n \sum_{\mu_l\nu_l=0}^1 e^{2 i (\mu_l-\nu_l)x_l}
= 4^n\prod_{l=1}^n \cos^2 x_l\:,
\ee
such that the average fidelity becomes
\ben\label{integral-fidelity}
F =  \int\! \fr{\diff{^n \b x}}{ N}
\:e^{-\b x\dag \b K\inv \b x }
\prod_{l=1}^n \cos^2 x_l \:.
\een
This relatively well-behaved integral representation of the
average fidelity can serve as starting point for numerical or
analytical calculations (cf.~Sec.~\ref{subsec-examples}). 
Furthermore, in contrast to the sum representation \Ref{sum-fidelity},
the integral \Ref{integral-fidelity} indicates how to obtain
approximative expressions.

\subsubsection{Weak coupling approximation}
\label{subsubsec-weak}
For weak couplings $\alpha$ the inverse eigenvalues of $\b K$
become large and hence the integrand sharply peaks at the
global maximum at $\b x=0$. In this case it is appropriate to expand the
integrand as 
\be
\:e^{-\b x\dag \b K\inv \b x }
\prod_{l=1}^n \cos^2 x_l
= \:e^{- \b x\dag (\b K\inv + \b 1) \b x \:+\: O(|\b
  x|^4)} 
\ee 
and to omit the $O(|\b x|^4)$ corrections in the exponent. Inserting
this in Eq.~\Ref{integral-fidelity} we arrive at a proper Gauss integral
which yields a surprisingly simple weak coupling approximation
\ben\label{det-fidelity}
F_{wc} = \det( \b 1 + \b K )^{-1/2}
\een
of the average fidelity. By consideration of a diagonal matrix $\b K$
we estimate the relative error $|F-F_{wc}|/F$ of order $\tr \b K^2$,
which one has to keep in mind when using this approximation.




\subsubsection{Small deviations}
\label{subsubsec-small}
Finally, let us consider the practically relevant situation where the
average fidelity deviates only by a small amount $\eps$ from unity.
By the weak coupling approximation we find
\be
F \equiv 1 - \eps = \det(\b 1 + \b K)^{-1/2} (1+ O(\tr \b K^2))\:,
\ee
and, taking the logarithms of both sides, 
\be
\ln (1-\eps) = -\fr{1}{2} \tr \ln( \b 1 + \b K) \: + \: O(\tr \b K^2)\:.
\ee
When we expand the logarithms we observe that in leading order
$
\eps = \fr{1}{2} \tr \b K\: + \: O(\tr \b K^2 ), 
$
and hence, since the decoherence matrix $\b K$ has constant diagonal
elements $K_0(t)$,
\ben\label{small-1-fidelity}
F = 1 \:-\: \fr{1}{2} n K_0(t)\: + \: O(\tr \b K^2 )\:.
\een
Small deviations of $F$ from unity are thus determined by the
zero-distance decoherence function $K_0(t)$, describing the decoherence of a
single (encoded) qubit, and they grow linearly with the number $n$ of qubits.


\subsection{Examples}\label{subsec-examples}
The following discussion of two illustrative examples will provide
more insights in the average register fidelity. The results will be
also useful in the subsequent comparison of plain and encoded quantum registers.

\subsubsection{Independent qubits}
\label{subsubsec-independent}
The first example is a quantum register consisting of independent
qubits, as it is reflected in vanishing decoherence functions $K_l(t)$
for $l>0$. For instance, this is realized in a plain, physical
register $R^0_n$ in the limit of a diverging lattice constant $a$, but
also holds to good approximation for an encoded register $R_n^1$ in
the long-time limit (cf. discussion in Sec.~\ref{subsec-discussion}).

As a consequence of $K_l(t)=0$ for $l>0$ the decoherence matrix $\b K$
of such a register is
\be
\b K = \kappa \: \b 1_n \:,
\ee 
where $\kappa = K_0(t)$, and $\b 1_n$ is the $n\times n$ unit matrix.
Because of this trivial matrix $\b K$ the integral in
\Ref{integral-fidelity} nicely factorizes
into $n$ one-dimensional Gaussian integrals, 
\ben\label{exact-uncorrelated}
F = \prod_{l=1}^n \int\!\fr{\diff{x_l}}{\sqrt{\pi \kappa}}
e^{-x^2_l/\kappa}\cos^2 x_l = 
\left( \fr{1+e^{-\kappa}}{2}
\right)^n\:. 
\een
Not unexpected, the average fidelity of the $n$-qubit register is
exactly the $n$th power of the average fidelity of a single qubit,
$F_1 =(1+e^{-\kappa})/2$. We note in passing that this result could
have been derived also directly from the sum representation
\Ref{sum-fidelity}. 

\begin{figure}
\vspace{0.5cm}
  \begin{center}
    \epsfxsize=7.0cm
    \epsffile{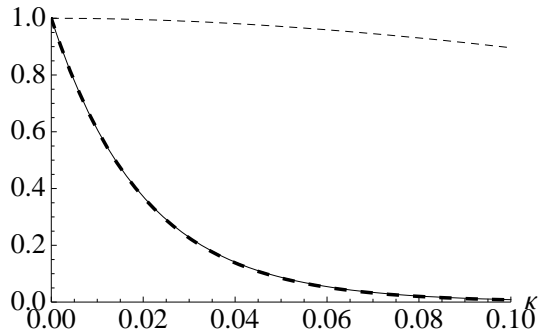}
    \vspace{0.3cm}
    \caption{Average fidelity of an 100-qubit register of independent
      qubits as a function of the single-qubit decoherence
      parameter $\kappa= K_0(t)$. The exact result (thick dashed curve) and the
      weak coupling approximation (solid curve) agree very well in the
      plotted regime. The upper dashed curve shows the ratio of exact and
      approximative fidelity.
    \label{fig-approximation}}
  \end{center}
\end{figure}

We can also employ the weak coupling expression \Ref{det-fidelity},
predicting 
\be
F_{wc} = (1+\kappa)^{-n/2}\:.
\ee
While this is not quite the exact result \Ref{exact-uncorrelated}, 
we indeed observe good agreement for small couplings $\kappa \ll
\sqrt{1/n}$.  
Notice that the weak coupling approximation particularly holds  in
the regime ${1}/{n} \ll \kappa \ll \sqrt{ {1}/{n} }$,
where the average fidelity is already exponentially small. 
(cf.~Fig.~\ref{fig-approximation}).

\subsubsection{Symmetrically coupled qubits}
\label{subsubsec-symmetrically}
As the extreme opposite to the first, our second example is a 
register whose qubits are symmetrically coupled to the bosonic bath by
an interaction Hamiltonian \Ref{symmetric-hamiltonian}. 
This is realized for a physical register in the limit of a vanishing
lattice constant $a$, where all qubits are located at the same
position. Consequently, here the decoherence matrix $\b K$ 
becomes a uniform matrix with constant entries
\ben\label{uniform_K}
\b K_{lm} = K_{|l-m|}(t) = K_0(t) \equiv \kappa\:. 
\een

Up to a factor $n\kappa$ the
matrix $\b K$ describes the orthogonal projection on the 
diagonal  $\b d = (1,\dots,1)/\sqrt{n}$. 
$\b K$ has therefore a non-degenerate eigenvalue $n \kappa$ with an
eigenvector $\b d$, and an $(n-1)$-fold degenerated eigenvalue $0$ with  
eigenspace $\b d^\perp$.
It follows that the integrand in \Ref{integral-fidelity}
has its entire weight on the diagonal $\b d$, as an effect of which
the $n$-dimensional integral collapses to a one-dimensional one.
In this way the average fidelity results in  
\bea
F &=& \fr{1}{\sqrt{\pi n \kappa}} \int \diff{x} \: e^{-x^2/n\kappa} \: \cos^{2n}
  \fr{x}{\sqrt{n}}  \\
&=& 
\fr{1}{4^n} \sum_{l=0}^{2n} \binomial{2n}{l} e^{- \kappa (n-l)^2}\:.
\eea
The result can be better interpreted in the limit of large $n\gg 1$ and 
small $\kappa \ll 1$ (independent of $n$). When $n$ is large, we can 
approximate the binomial factor $4^{-n} \binomial{2n}{l}$ by a Gaussian, 
$ \exp(- (n-l)^2/n)/\sqrt{\pi n}$, and further, when $\kappa$ is small, we
are allowed to replace the sum by an integral. This  yields an
average fidelity
\ben\label{large_n_fidelity}
F = \fr{1}{\sqrt{1+n \kappa}}\:.
\een
We notice that this expression also results from the weak coupling
approximation Eq.~\Ref{det-fidelity}, since here $\det(\b 1 ~  + ~ \b K) ~ = ~ 1 ~ + ~ n \kappa$. 

The algebraical decay with $n$ in Eq.~\Ref{large_n_fidelity} strongly contrasts with the exponential decay of the average register fidelity Eq.~\Ref{exact-uncorrelated} observed for independent qubits. 
This marked difference must be attributed to the high degree of symmetry
in the present case. In fact, the symmetric qubit-boson coupling
\Ref{symmetric-hamiltonian} 
entails that states with a small total spin-$z$ component couple much
less effectively to the bosonic bath as than they would do in the case
of independent qubits. 
Apparently, for numbers of qubits and couplings with $n \gg 1/\kappa$ this results in a strongly enhanced averaged fidelity. 

\begin{figure}
\vspace{0.5cm}
  \begin{center}
    \epsfxsize=7.0cm
    \epsffile{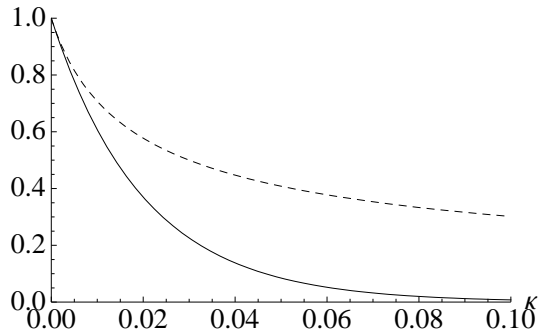}
    \vspace{0.3cm}
    \caption{Average fidelities of an 100-qubit register for
      independent (solid curve) and symmetrically coupled (dashed curve) qubits
      as a function of the decoherence parameter $\kappa=K_0(t)$. 
    \label{fig-approximation2}}
  \end{center}
\end{figure}

Remarkably, a register with $n \ll 1/\kappa$ does not benefit from these effects of symmetry. 
In this regime, the average fidelity for independent and symmetrically coupled qubits actually coincide (cf.~Fig.~\ref{fig-approximation2}).
We are lacking a simple
explanation for that, however, since in this regime also $1-F \ll 1$
we can refer to the general result Eq.~\Ref{small-1-fidelity} for
small deviations. According to this relation, here the fidelity is
dominated solely by the zero-distance decoherence function $K_0(t)$,
and hence all details concerning the spatial structure  of the
register do not matter.  

\subsection{Comparison of plain and encoded quantum register}
\label{subsec-comparison-register}
In this subsection we will compare a plain, physical register $R_n^0$
with a $1^{st}$ order encoded register $R_n^1$ by means of their
respectice average register fidelities $F_0$ and $F_1$. Thereby, we
will make good use 
of the results for the two preceding examples.
We will restrict the comparison to the high temperature regime $a\gg
1/T$, where we can use the relatively simple expressions
Eqs.~\Ref{Knull-large}, 
\Ref{largeT1}, and \Ref{largeT2} for the decoherence function.  
Furthermore, the time $t$ will be assumed to be larger than $L_0=(n-1)a$
for the plain register $R_1^0$ and larger than $L_1 = (2n-1)a$ for the
encoded register $R_n^1$.

\subsubsection{Average fidelity of a plain quantum register}
We consider a physical $n$-qubit register ($R_n^0$) as defined in
Sec.~\ref{subsec-plain}. 
In the high temperature regime and for $t ~ > ~ L_1$, its time dependent
decoherence matrix 
$\b K(t)$ follows by Eqs.~\Ref{Knull-large} and \Ref{largeT1} to 
be given by
\ben\label{linear-K}
\b K_{lm}(t) = \gamma t \left(1\: - \: \fr{a|l-m|}{2 t} \right)\:,\quad
\gamma = \alpha \pi T\:, 
\een
where we suppressed a logarithmic term $ \alpha \ln( \Omega/2\pi T) ~ \ll ~ \gamma t$ in the diagonal matrix elements.
Since the non-trivial structure of $\b K(t)$ does not allow for a simple
evaluation of the exact formula Eq.~\Ref{integral-fidelity}, 
we immediately switch to a numerical evaluation of the weak coupling
approximation Eq.~\Ref{det-fidelity}, 
\be 
F_0(t) = \det(\:\b 1 + \b K(t)\:)^{-1/2}\:.
\ee

\begin{figure}
\vspace{0.5cm}
  \begin{center}
    \epsfxsize=7.0cm
    \epsffile{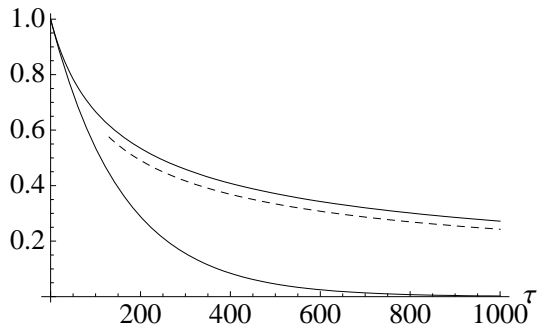}
    \vspace{0.3cm}
    \caption{Time dependency of the average fidelity for 
    a linear quantum register of 125 qubits (dashed curve) in comparison with the 
    average fidelity of registers consisting of 125 independent
    (bottom) and symmetrically (top) coupled qubits as a function of dimensionless time $\tau = t / a$. 
		The rate is $\gamma = 10^{-4} / a$. 
    \label{fig-foft}}
  \end{center}
\end{figure}

The dashed curve in Fig.~\ref{fig-foft} shows the average fidelity $F_0(t)$ 
of a linear qubit register with $n=125$ qubits at a decoherence
rate $\gamma = 10^{-4} /a$. The time domain $130 a < t < 1000 a$ is
chosen such that Eq.~\Ref{linear-K} and the weak coupling
approximation is applicable. 
For comparison, Fig.~\ref{fig-foft} also shows the average register 
fidelities of independent 
and symmetrically coupled 
qubits with a time dependent parameter $\kappa= \gamma t$, 
\bean
F_i(t) &=& \left(\fr{1+e^{-\gamma t}}{2} \right)^n\:,  \nn \\
F_s(t) &=& \fr{1}{\sqrt{1 + n \gamma t}} \label{F_s}
\eean
(cf.~Eqs.~\Ref{exact-uncorrelated} and \Ref{large_n_fidelity}).

As expected, the average register fidelity $F_0(t)$ lies
between $F_i(t)$ and $F_s(t)$. 
It might be more surprising that for large times  $t \gg  L_0$ the fidelity
$F_0(t)$ is much closer to the fidelity $F_s(t)$ of a
register of symmetrically coupled qubits than to $F_i(t)$, the
fidelity corresponding to independent qubits.
The reason for this resemblance of $F_0(t)$ and $F_s(t)$ 
is 
that for $t \gg  L_0$ the $l, m$-dependent term in Eq.~\Ref{linear-K}
is a small correction to the leading term, meaning that $\b K$ becomes
close to the uniform decoherence matrix discussed in
Sec.~\ref{subsubsec-symmetrically} with $\kappa = \gamma t$.   



At small times $t <  L_0$, which are not covered by the decoherence
matrix Eq.~\Ref{linear-K}, 
relation Eq.~\Ref{small-1-fidelity} predicts 
the fidelity $F_0(t)$ to agree with the converging fidelities for
independent and symmetrically coupled qubits.

\subsubsection{Average fidelity of plain and encoded registers}
\label{subsubsec-encoded}
To begin with, we consider the average fidelity $F_1$ of an encoded
register $R_n^1$ at large temperatures $T \gg 1/a$. For times $t> L_1$ we find
from relation Eq.~\Ref{chi-decoherence-function} with Eqs.~\Ref{Knull-large},
and \Ref{largeT1} a decoherence matrix
\bea
\b K_{ll}(t)\quad &=& \: K_0^1(t) \: = \: \gamma a\:, \quad \gamma =
\alpha \pi T\:, \\
\b K_{l\neq m}(t) &=& \: 0\:,
\eea
where again we  omitted a logarithmic term $\alpha \ln(\Omega/2\pi T)
\ll \gamma a$ in the diagonal elements. Note that the time dependence
has dropped out.
This is precisely the decoherence matrix of a register with
independent qubits which we analyzed in
Sec.~\ref{subsubsec-independent}. Hence,
by Eq.~\Ref{exact-uncorrelated}, for times $t>L_1$
the average fidelity becomes a constant
\ben\label{F_1}
F_1(\infty) = F_1 = \left( \fr{1 + e^{- \gamma a}}{2} \right)^n\:.
\een
For instance, for $\gamma = 10^{-4}/a$ and $n=125$, which are the parameters of the
fidelity $F_0(t)$ plotted in Fig.~\Ref{fig-foft}, we obtain an $F_0(t)$
largely exceeding asymptotic value
\be
F_1(\infty) \simeq 1 - 0.0062 \:.
\ee
%
%

The saturation of the effective decoherence functions $K_l^1(t)$
at times $t > a$ entails the saturation of the averaged register fidelity
$F_1(t)$ to the asymptotic value $F_1(\infty)$. In contrast to that,
the fidelity $F_0(t)$ of the plain quantum register keeps decaying
with increasing time. As a trivial consequence, the average fidelity
of the encoded register will always exceed the fidelity of a plain register
if time $t$ becomes sufficiently large.

To resolve the time dependency of the average fidelities at shorter times
we may use approximation Eq.~\Ref{small-1-fidelity}, valid for small
deviations
$1-F\ll 1$, according to which
\be
F_0(t) = 1 - \fr{n}{2} K_0(t)\:, \quad
F_1(t) = 1 - \fr{n}{2} K_0^1(t)\:,
\ee
where $K_0(t)$ and $K_0^1(t)$ are decoherence functions for plain and
encoded qubits, respectively.
We therefore expect encoding to be advantageous when $K_0(t) >
K_01(t)$. This is the case for times $t$ larger than the crossover time
$t_c$ which in Sec.~\ref{subsec-discussion} was determined to be $t_c
\simeq a$ for large temperatures $T\gg 1/a$, and
$t_c \simeq \fr{1}{\pi T} \ln(2 \pi a^2 \Omega T/e^2) $ for low
temperatures $T \ll 1/a$.

\begin{figure}[h]
\vspace{0.5cm}
  \begin{center}
    \epsfxsize=7.0cm
    \epsffile{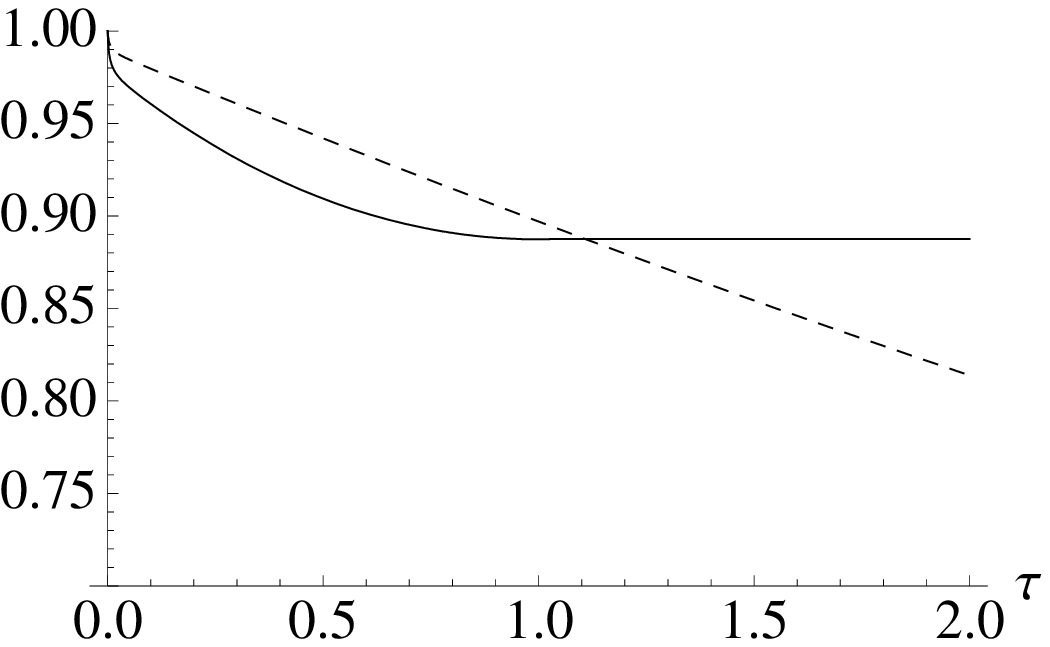}
    \epsfxsize=7.0cm
    \epsffile{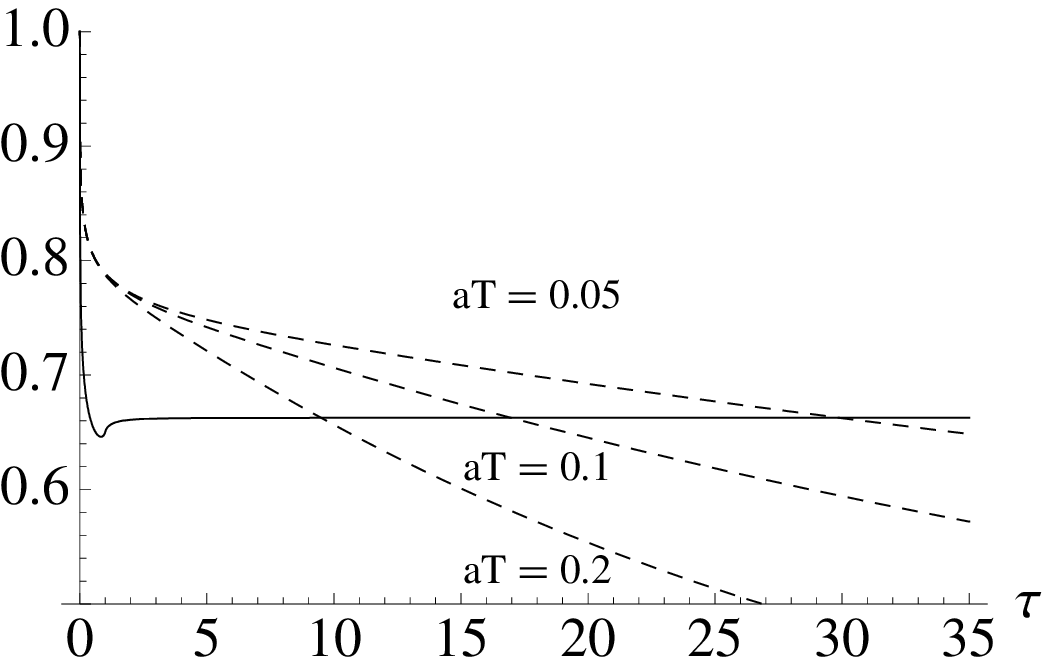}
    \vspace{0.3cm}
    \caption{
			Average fidelity of encoded (solid line) and plain registers (dashed line) of 125 qubits for a large temperature $T ~ = ~ 10 / a$ in the upper graph and for low temperatures $a T ~ = ~ 1/5 , 1/10, 1/20$ in the lower graph as function of the dimensionless time $\tau = \nicefrac{t}{a}$. The coupling is $\alpha = 10^{-3} / a$. Notice that at low temperatures the fidelity of the encoded register is almost temperature independent.
			\label{fig-temperatur-fidelity}
		}
  \end{center}
\end{figure}

We confirmed this by numerical calculation of the fidelities
$F_0(t)$ and $F_1(t)$ within the weak coupling approximation
Eq.~\Ref{det-fidelity}. 
Fig.~\ref{fig-temperatur-fidelity} shows the fidelity of encoded and plain registers in two temperature regimes ($T \gg 1/a$ and $T \ll 1/a$). The occuring crossover times are in good agreement with the above discussed expectations.

Finally, it might be instructive to analyze the requirements on
the single-spin decoherence rate $\gamma = \alpha \pi T$ which is
needed in order to store quantum information within a given precision $\eps$
during a given time period $t_s$. (For simplicity, we restrict ourself
to large temperatures $T\gg 1/a$.) Let us assume that $t_s$ scales with
the number of qubits as
\be
t_s \simeq  t_0 n^q\:,
\ee
where $q$ is some power greater or equal unity, and $t_0 \sim a $ is
some microscopic time scale. Then, for the plain register the
condition
\be
F_0(t_s) \ge 1-\eps
\ee
implies either by Eq.~\Ref{small-1-fidelity} or by $F_0(t) \approx
F_s(t)$ and Eq.~\Ref{F_s} a rate $\gamma_0$ scaling as
\be
\gamma_0 \sim \fr{2\eps}{t_0} n^{-(1+q)}\:.
\ee
On the other hand, for the encoded register the condition $F_1 \ge
1-\eps$ is satisfied if, according to Eq.~\Ref{F_1}, the rate scales as
\be
\gamma_1  \sim  \fr{2\eps}{a} n^{-1}\:,
\ee
The advantage of using an encoded register over using a plain
register is reflected in a possibly huge factor
\be
\fr{\gamma_1}{\gamma_0} \sim \fr{t_0}{a} n^q\:.
\ee 

\section{Dissipative couplings}\label{sec-dissipative}
The preceding sections have shown that the decoherence of an encoded qubit and also the fidelity of an encoded register (of $1^{st}$ order) is essentially determined by the effective decoherence function
\be
K_0^{1}(t) = 2(K(0,t) - K(a,t)) .
\ee
In accordance with previous work\cite{palma97,doll07} we observed that for the dissipation-less model the decoherence coefficient $K_0^{1}(t) $ saturates to a finite constant value $K_0^{1}(\infty)$ in the long time limit. 
It is important to find out whether this saturation holds for general physical interactions, or merely is a feature of the dissipation-less interaction. 

Here we cannot address this question in full generality.
However, as a first step we will analyze a two-spin system that is dissipatively coupled to a bosonic environment. 
Lacking an exact analytical solution, we will treat this system by a quantum master-equation of Bloch-Redfield type (Sec.~\ref{subsec-masterequation}). 
Its viability in the present context is demonstrated for the dissipationless case, where the exact results can be reproduced (Sec.~\ref{subsec-dissipation-less}).
In Sec.~\ref{subsec-dissipative} we will then consider the decay of a state that would remain invariant under the spin-boson interaction in the limit $a \to 0$.
Contrary to what is observed in the dissipation-less case, here we find that for any finite $a$ the state continues to decay with a constant rate for large times. 
The asymptotic rate appears to be the given by the decay rate of a single spin multiplied with a universal factor $2-2\sin(\eps a)/(\eps a)$.

\subsection{Bloch-Redfield master equation}\label{subsec-masterequation} 
We consider the general situation of a system $S$ that is coupled to a bath $B$ via an interaction $H_I$.
The Lioville-von-Neumann equation for the density operator $\rho_{SB}$ of the total system in the interaction picture is
\be
\dot\rho_{SB}(t) = -i [ H_I(t), \rho_{SB}(t)] .
\ee
It follows that the reduced density operator $\rho$ (interaction picture) of the system obeys an equation of motion\cite{breuer02}
\begin{eqnarray}
	\label{exact_eom}
	\dot\rho(t)
	& = & - \int_0^t \diff{s} \tr_B [ H_I(t), [H_I(s),\rho_{SB}(s)]]\\
	&   & - i \tr_B [H_I(t), \rho_{SB}(0) ].
\end{eqnarray}
We assume that initially $\rho_{SB}(0)$ is a product of an initial $\rho(0)$ and a thermal bath state $\rho_B$, and further take for granted that 
\ben\label{zero_commutator}
\tr_B [H_I(t),\rho(0) \otimes \rho_B  ] =0 ,
\een
which in many cases is satisfied, in particular in those to be analyzed below.
The remaining term on the r.h.s. depends on the total state $\rho_{SB}(s)$ at times $0 \le s \le t$. In general, this does no allow to exactly determine $\rho(0)$ from Eq. \Ref{exact_eom}. 
One therefore frequently invokes the Born-Markov approximation by substituting
\ben\label{born-markov}
\rho_{SB}(s) \to   \rho(s) \otimes \rho_B  \to  \rho(t) \otimes
\rho_B,
\een
Obviously, this approximation is good in the limit of weak couplings.
This results in the Bloch-Redfield master equation 
\ben\label{redfield-equation}
\dot\rho(t) = \cl R_t(\rho(t)),
\een
where $\cl R_t$ denotes the Redfield super-operator defined by 
\ben\label{redfield-operator} 
\rho \mapsto R_t(\rho) = - \int_0^t \diff{s} \tr_B [H_I(t), [H_I(s), \rho \otimes
\rho_B]].
\een

Note that the Redfield operator is explicitly time-dependent and
therefore the resulting dynamics does not exhibit a semigroup
structure. 
In this sense, the Bloch-Redfield equation Eq. \Ref{redfield-equation}
is non-Markovian, notwithstanding the fact that the Born-Markov
approximation has been used to derive it. 
In many cases it is justified to eliminate this ``deficiency'' by simply extending the domain of integration in Eq.~\Ref{redfield-operator} from $[0,t]$ to $[-\infty,t]$ (cf.~Ref.~\lite{breuer02}).  
However, as it has been stressed by Doll et al.\cite{doll07}, when
dealing with a spatially extended quantum object this procedure would
lead to noncausal behavior and thus to spurious results.  

In Ref.~\lite{doll07} this problem has been circumvented by using a {\em
  causal} master equation in which causality is explicitly taken care
of by step functions in the time domain that truncate acausal
contributions.   
The resulting dynamics has been shown to approximate quite well the known exact solution.  
Here, we will simply stay with the non-Markovian Bloch-Redfield
equation as given by Eqs. \Ref{redfield-equation} and
\Ref{redfield-operator}. 

\subsection{Dissipation-less two-spin system}\label{subsec-dissipation-less} 

First, in order to demonstrate its viability, we use the
Bloch-Redfield master equation to reanalyze the dissipation-less model
of Section \Ref{subsec-plain} for $n=2$.  
The spin-boson Hamiltonian in the interaction picture is
\bean
\label{hamiltonian_a}
H_I(t) = \sum_{l=0,1} Z_l \otimes B({\b r}_l,t),
\eean
where $Z_l$ is the (time-independent) Pauli-$z$-operator on the $l$-th spin, and $B({\b r}_l,t)$ as in Eq.~\Ref{bosonic-field}.
Condition \Ref{zero_commutator} is satisfied and the Redfield operator determines to be
\be
\cl R_t(\rho) = \sum_{m,l = 0,1} C(|{\b r}_l-{\b r}_m|,t) \left( Z_m \rho Z_l - Z_l Z_m \rho \right)  +  h.c.,
\ee
where 
\be
C(|{\b r}|,t) = \int\limits_0^{t} \diff{t'} \sp{ B(\b r, t')B(\b 0,0) }_T\: .
\ee
Presenting $\rho(t)$ in the computational basis $\ket \mu$, 
\be
\rho(t) = \sum_{\mu,\nu} \rho_{\mu\nu}(t) \ket{\mu}\bra{\nu},
\ee
and again omitting imaginary parts, which would contribute only to the unitary pure $\cl U$ of the time evolution, the Bloch-Redfield master equation \Ref{redfield-equation} predicts the coefficients $\rho_{\mu\nu}(t)$ to obey independent differential equations  
\bea
\dot \rho_{\mu\nu}(t) &=& \! \sum_{m,l=0}^1 (\mu_l \! - \! \nu_l)(\mu_m \! - \! \nu_m)  4 \Re C(a\!|m\!-\!l|,t) \! \rho_{\mu\nu}(t). 
\eea
After integration we observe that the master master-equation reproduces the exact result Eqs.~\Ref{decoherence-coefficients} and \Ref{correlator-integral} obtained in Sec.~\ref{sec-decoherence} (for $n=2$).
This is more than one could have expected and we believe that the exactness must be ascribed to the fact that the present model is lacking dissipative couplings. 
We do not expect that the Bloch-Redfield theory exactly describes the dynamics in the dissipative model. 
Nevertheless, the positive outcomes for the present dissipation-less model still encourages us to use the Bloch-Redfield master equation also for the dissipative model that we are investigating next.  


\subsection{Dissipative two-spin system}\label{subsec-dissipative}
Let now the two spins interact with the bosonic bath via the
dissipative Hamiltonian (interaction picture) 
\bean 
\label{hamiltonian_b}
H_I(t) = \sum_{l=0,1} X_l(t) \otimes B({\b r}_l,t).
\eean
where $B({\bf r},t)$ is as in Eq. \Ref{bosonic-field}. $X_l(t)$ can be conveniently written with operators
\bea
u_l & \equiv &  \sigma_+^{(l)} = X_l + i Y_l \\
d_l & \equiv &  \sigma_-^{(l)} = X_l - i Y_l 
\eea
as
\be
X_l(t) = e^{-i \eps t} d_l + e^{+i \eps t} u_l.
\ee

\subsubsection{Master equation}
A straightforward calculation shows that in rotating wave
approximation -- which is valid as long as $\eps t \gg 1$ -- the
corresponding Redfield operator Eq. \Ref{redfield-operator} is given
by   
\bean
\cl R_t(\rho) &=& \sum_{m,l}
 C_{-}(|{\b r}_l-{\b r}_m|,t) \left\{ u_m \rho d_l - d_l
 u_m  \rho  \right\} \nn\\
& & \quad +
 C_{+}(|{\b r}_l-{\b r}_m|,t) \left\{ d_m \rho u_l - u_l
 d_m  \rho  \right\} \nn\\
& & \quad + h.c., \label{dissipative-redfield}
\eean
with energy dependent correlation functions 
\be
C_{\pm}(|{\b r}|,t) = \int_0^t \diff{s}  e^{ \pm i \eps s} \sp{B({\b r},s) B(0,0)}.
\ee
The real part $C'_\eps(r,t)$ of $C_\eps(r,t)$ has a surprisingly
simple structure in the limit of large $Tt \gg 1$. 
Here we find the zero-distance correlations to be given by the familiar expressions\cite{breuer02}
\bean
C'_{-}(0,t) &=& \alpha \pi  n(\eps) \nn\\
C'_{+}(0,t) &=& \alpha \pi  \left( n(\eps)+1 \right),
\label{pm-zero-correlations} 
\eean
where $n(\eps) = (e^{\eps / T}-1)^{-1}$.
The finite-distance correlations follow to be ($|{\bf r}| = a$)
\ben\label{pm-correlations}
C'_{\pm}(a,t) = \fr{\sin (\eps a)}{\eps a}  \theta(t-a)  C'_{\pm}(0,t).
\een
We notice that the finite-distance correlation deviates from the zero-distance correlation only by a temperature-independent factor that, of course, approaches unity for $a \to 0$, but also for $\eps \to 0$ (when $a<t$), what is somewhat unexpected.  

Anticipating the discussion given below, we remark that the deviation of the first factor $\fr{\sin (\eps a)}{\eps a}$ from unity for any finite $\eps r$ reflects the decay of coherence in a symmetric subspace in the long time limit.
Note that for $\eps \to 0$, which precisely corresponds to the transition to a dissipation-less model, the first factor remains unity also for finite $a$.
Clearly, this corresponds to the saturation of the subspace fidelity observed in the previous section.  
The second factor, $\theta(t-a)$, which is responsible for causality, here as well as in the previous case leads to an initial drop of the subspace fidelity until the time $t=a$ is reached. 

\subsubsection{Asymptotic decay of subspace fidelity}
The anti-symmetric state
\be
\ket{\psi_0} = \fr{1}{\sqrt 2} \left( \ket{01} - \ket{10} \right).
\ee
is annihilated by $Z_0 + Z_1$ and $X_0 + X_1$,
and therefore remains invariant under both, the dissipationless interaction Eq.~\Ref{hamiltonian_a} and the dissipative interaction Eq.~\Ref{hamiltonian_b},
provided that $a=0$.
We are interested in the decay of the state $\ket{\psi_0}$ under the dissipative spin-boson interaction Eq.~\Ref{hamiltonian_b} for finite distance $a$.
To this end we consider the fidelity
\be
F(t) := \sp{ \psi_0 | \rho_0(t) | \psi_0},
\ee
where $\rho_0(t)$ is the reduced spin state at time $t$ (in interaction picture) that evolved via the interaction Eq.~\Ref{hamiltonian_b} with the bath from the initial state $\rho(0)= \pro{ \psi_0 }$. 
Here we will approximately determine $\rho_0(t)$ by the Bloch-Redfield master equation with the dissipative Redfield operator Eq. \Ref{dissipative-redfield}. 

\begin{figure}
\vspace{0.5cm}
  \begin{center}
		\epsfxsize=7.0cm
    \epsffile{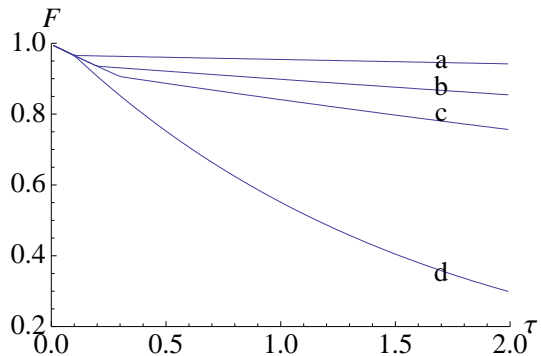}
    \vspace{0.3cm}
    \caption{
			Fidelity of a symmetric subspace in a dissipative model as a function of dimensionless time $\tau = t T$ with $\eps = 5 T$ for inter-spin distances $a=0.1/T (a),0.2/T (b)$ and $0.3/T (c)$.
			The larger the distance, the larger is the asymptotic decay rate at $t > a$. 
      For comparison, the lower curve (d) shows the fidelity for the symmetric state $\fr{1}{\sqrt 2} \left( \ket{01} + \ket{10} \right)$.     
    \label{fig-fidelity}}
  \end{center}
\end{figure}

To obtain a first impression of the dynamics we integrated the master equation numerically.
Characteristic outcomes for the fidelity $F(t)$ are shown in Fig. \ref{fig-fidelity}.
We chose energy $\eps = 5 T$ and a small overall coupling constant $\alpha = 0.01$. 
Similar to the previously observed behavior, also here we see a  relatively strong decay of the fidelity at times $t < a$. 
However, in contrast to the dissipation-less model, here we clearly see a decay of $F_0(t)$ for large times with a rate that increases with distances ranging from $aT=0.1$ to $0.3$. 

\begin{figure}
	\vspace{0.5cm}
  \begin{center}
  	\epsfxsize=7.0cm
    \epsffile{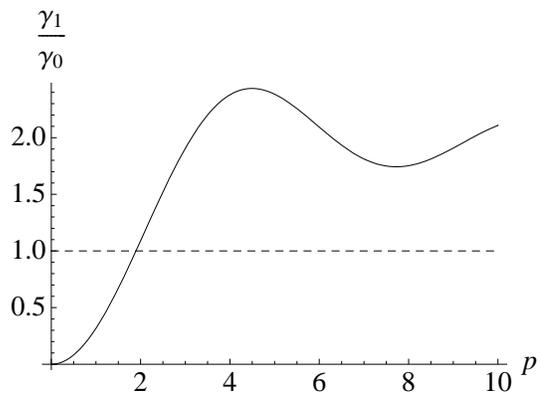}    
    \vspace{0.3cm}
    \caption{The reduction $\gamma_1/ \gamma_0$ in the decay rate as a
      function of the dimensionless parameter $p= \eps a$.
    \label{fig-reduction}}
  \end{center}
\end{figure}

At times $t \gg a$ and for sufficiently small couplings the decay rate $\gamma_1 = \diff{F} / \diff{t}$ of the fidelity $F(t)$ is in good approximation determined by the expression
\be
\gamma_1 = - \sp{\psi_0 | \cl R_t( \pro{\psi_0}) | \psi_0} .
\ee
Using expressions Eqs.~\Ref{pm-zero-correlations} and \Ref{pm-correlations} we obtain after some algebra
\be
\gamma_1 =  2\left( 1 - \fr{ \sin (\eps a)}{\eps a} \right) \gamma_0
\ee
where $\gamma_0 = 2 \pi \alpha (n(\eps) + 1/2)$ is the decay rate of a single spin.   
This is a simple and quite general result that -- in the light of the discussions of the preceding sections -- quantifies the benefits of using a symmetric subspaces in a dissipative system. 
It identifies 
\be
p = \eps a \equiv \fr{ \eps a}{ \hbar c}
\ee
as the relevant parameter that captures the achievable reduction of the decay rates of encoded qubits in comparison to plain physical qubits. 
Fig.~\ref{fig-reduction} shows $\gamma_1/ \gamma_0$ as a function of $p$.

A significant reduction requires $p \ll 1$, corresponding to distances
\be
a \ll \fr{c \hbar}{\eps}.
\ee
For instance, for atomic qubits this implies that the distance $a$ between atoms should be small compared to the wavelength of the light that is emitted in a bit-flip transition. 

\section{Summary of results}\label{sec-summary}

In Secs.~\ref{sec-decoherence} and \ref{sec-fidelity} we studied a
quantum register physically realized as a linear spin-chain that
interacts with a three-dimensional bosonic bath via a
dissipation-less spin-boson coupling  with an ohmic coupling spectral
density. Within this framework we analyzed the benefits of using
encoded qubits and encoded quantum registers in order to reduce effects of decoherence. 

In agreement with previous work \cite{palma97,doll07} we found that the coherence of a $1^{st}$ order encoded qubit converges to a finite asymptotic value at times $a/c$, where $a$ is the inter-spin distance and $c$ is the velocity of sound or light. 
As a consequence, the coherence of the
encoded qubit exceeds the one of a plain qubit for times larger than a
crossover time $t_c$. At high temperatures $k_B T \gg \hbar c /a$ the
crossover time is $t_c \simeq a/c$, whereas at low temperatures $k_B T
\ll \hbar c/a$ it increases roughly as $t_c \simeq \hbar /(k_B T)$
with decreasing temperature $T$. 
Moreover, we observe the $1^{st}$ order
encoded qubits to be effectively decoupled from each other, meaning that
higher-order encoding becomes counterproductive.  On the other hand,
this decoupling of encoded qubits should be advantageous for quantum
error correction, which is known to be significantly hampered when the
qubits are coupled via the bosonic bath \cite{klesse05}. This aspect
of using encoded qubits may deserve further investigation. 

In Sec.~\ref{sec-fidelity} we derived a convenient integral
representation for the averaged register fidelity with respect to the
noise of the dissipation-less spin-boson model. This allowed us to
investigate the decoherence of an entire plain or encoded quantum
register. Small deviations of the averaged register fidelity from
unity are proportional to the number of (encoded) qubits and to the
(effective) decoherence function $K_0(t)$ of a single (encoded)
qubit. The improved performance of $1^{st}$ order encoded qubits therefore carries
over to an entire $1^{st}$ order encoded quantum register. This is
confirmed by more detailed analytical and numerical results presented
in Sec.~\ref{sec-fidelity}. 

Finally, in Sec.~\ref{sec-dissipative} we addressed the role of
dissipation within a two-spin model. Its dynamics was determined by
employing a master equation of Bloch-Redfield type. 
In the presence of
dissipative spin-boson couplings the coherence of an encoded qubit seems
no longer to converge to a finite value. Instead, here we expect an
exponential decay with an, however, reduced asymptotic effective decay
rate $\gamma_{1} = 2(1-\sin(p)/p)\gamma_0$, where the dimensionless
parameter $p$ is determined by the energy splitting $\eps$ of the
spins and the inter-spin distance $a$, $p= \eps a / (\hbar c)$. 

This work is supported by DFG grant No. KL2159.

\begin{appendix}

\section{Decoherence of an encoded quantum register}\label{app-decoherence-of-encoded}  
To begin with, let us consider a $1^{st}$ order encoded register
$R_n^1$. Its Hilbert space $H_n^1$ is spanned by $2^n$ state vectors 
\be
\ket{\mu}^1 = \ket{\mu_0}_0^1 \dots \ket{\mu_{n-1}}_{n-1}^1 \in
H_{2n}\:, \quad \mu \in \b Z_2^n\:.
\ee
Each $\mu \in \b Z_2^n$ corresponds one-to-one to a $\mu' \in \b Z_2^{2n}$ by
demanding 
$
\ket{ \mu }^1 = \ket{\mu'}\:,
$
which by definition \Ref{logical-kets} means 
\ben\label{mu-relation}
\mu'_{2i}=\mu_i\:, \quad
\mu'_{2i+1} = 1-\mu_i\:.
\een
Let $\rho_1$ be a state of the encoded register, i.e. $\mbox{supp}\:\rho_1 ~ \subset ~ H_n^1$.
By Eq.~\Ref{N-operation} we find
\be
\cl N(\rho_1) = \sum_{\mu\nu \in \b Z_2^n} e^{D_{\mu'\nu'}}
  \ket{\mu}^{\!1}\!\bra{\mu} \rho_1 \ket{\nu}^{\!1}\!\bra{\mu}\:,
\ee
where $\nu'$ relates to $\nu$ as $\mu'$ to $\mu$, and, by
Eq.~\Ref{decoherence-coefficients},
\be
D_{\mu'\nu'} = \sum_{lm=0}^{2n-1} (\mu'_l -\nu'_l)(\mu'_m-\nu'_m) K( |l-m|a,t)\:.
\ee 
Making use of relation \Ref{mu-relation} and rearranging
the sum we can rewrite $D_{\mu'\nu'}$ in terms of $\mu$ and $\nu$ as
an effective decoherence coefficient
\ben\label{effective-decoherence}
D^1_{\mu\nu} = \sum_{lm=0}^{n-1}  (\mu_l -\nu_l)(\mu_m-\nu_m) K_{|l-m|}^1(t)\:,
\een
when effective decoherence functions $K^1_{l,m}(t)$ are defined as 
\be
K_{l}^1(t) = 2 K_{2l}^0(t)-K_{|2l-1|}^0(t) - K_{2l+1}^0(t)\:, 
\ee
with 
\be
K^0_{l}(t) = K(la, t)\:.
\ee

It is straightforward to generalize this analysis to higher orders,
which eventually leads us to relations \Ref{chi-N-operation},
\Ref{chi-decoherence}, and  \Ref{chi-decoherence}. 

\end{appendix}

\bibliography{paper200704}


\def \pra#1#2#3#4{ #1, Phys.~Rev.~A {\bf #2}, #3 (#4)} 
\def \prb#1#2#3#4{ #1, Phys.~Rev.~B {\bf #2}, #3 (#4)} 
\def \prl#1#2#3#4{ #1, Phys.~Rev.~Lett. {\bf #2}, #3 (#4)} 
\def \nat#1#2#3#4{ #1, Nature {\bf #2}, #3 (#4)} 
\def \sci#1#2#3#4{ #1, Science {\bf #2},#3 (#4)}


\end{document}